\documentclass[showpacs,floatfix,superscriptaddress,showpacs,twocolumn,amssymb,amsfonts,prb,aps]{revtex4}
\usepackage{longtable,graphicx,epsfig,dcolumn}

\begin{document}
\bibliographystyle{revtex}
\newlength{\defbaselineskip}
\setlength{\defbaselineskip}{\baselineskip}
\newcommand{\setlinespacing}[1]%
           {\setlength{\baselineskip}{#1 \defbaselineskip}}
\newcommand{\doublespacing}{\setlength{\baselineskip}%
                           {2.0 \defbaselineskip}}
\newcommand{\singlespacing}{\setlength{\baselineskip}{\defbaselineskip}}

\title{Surface Structure of Liquid Metals and the Effect of Capillary
Waves: X-ray Studies on Liquid Indium}

\author{Holger Tostmann}
\affiliation{Division of Engineering and Applied Sciences, Harvard
University, Cambridge Massachusetts 02138 (USA)}

\author{Oleg~Shpyrko}
\affiliation{Department of Physics, Harvard University, Cambridge
Massachusetts 02138 (USA)}

\author{Peter~S.~Pershan}
\affiliation{Department of Physics, Harvard University, Cambridge
Massachusetts 02138 (USA)} \affiliation{Division of Engineering
and Applied Sciences, Harvard University, Cambridge Massachusetts
02138 (USA)}

\author{Elaine~Dimasi}
\affiliation{Department of Physics, Brookhaven National Lab, Upton
New York 11973 (USA)}

\author{Ben~Ocko}
\affiliation{Department of Physics, Brookhaven National Lab, Upton
New York 11973 (USA)}

\author{Moshe~Deutsch}
\affiliation{Department of Physics, Bar-Ilan University, Ramat-Gan
52900 (Israel)}

\date{30 July 1998}

\begin{abstract}
We report x-ray reflectivity (XR) and small angle off-specular
diffuse scattering (DS) measurements from the surface of liquid
Indium close to its melting point of $156^\circ$C. From the XR
measurements we extract the surface structure factor convolved
with fluctuations in the height of the liquid surface. We present
a model to describe DS that takes into account the surface
structure factor, thermally excited capillary waves and the
experimental resolution. The experimentally determined DS follows
this model with no adjustable parameters, allowing the surface
structure factor to be deconvolved from the thermally excited
height fluctuations. The resulting local electron density profile
displays exponentially decaying surface induced layering similar
to that previously reported for Ga and Hg. We compare the details
of the local electron density profiles of liquid In, which is a
nearly free electron metal, and liquid Ga, which is considerably
more covalent and shows directional bonding in the melt. The
oscillatory density profiles have comparable amplitudes in both
metals, but surface layering decays over a length scale of $3.5\pm
0.6$~\AA\ for In and $5.5\pm 0.4$~\AA\ for Ga. Upon controlled
exposure to oxygen, no oxide monolayer is formed on the liquid In
surface, unlike the passivating film formed on liquid Gallium.
\end{abstract}


\maketitle

\section{Introduction}

The structure of the free surface of a liquid metal is fundamentally
different from that of a dielectric liquid such as water or Argon.
For dielectric liquids, the interatomic or intermolecular potential
is long-ranged and the nature of the interactions does not change
significantly across the liquid--vapor interface. Theoretical modeling
and molecular dynamics studies result in an interfacial density profile
that varies monotonically from the bulk liquid density to the bulk
vapor density.\cite{abra82} This can be contrasted with results of
computer simulations of dielectric liquid noble gases in contact with
their solid phase.\cite{chap77} There, the hard wall provided by the
solid phase is shown to induce surface layering in the liquid with a
well defined lamellar structure.

In liquid metals, the potential energy function changes
drastically
from a screened short-ranged Coulomb potential in the
bulk liquid phase where the conduction electrons are delocalized
 to a long-ranged van der~Waals type potential in the
vapor phase where all electrons are localized.\cite{rice74}
In this case, density functional calculations\cite
{evan81,iwam92} and computer simulations\cite{harr87} indicate
an oscillatory surface-normal density profile.
This surface induced layering can be attributed to the fact that strong Coulomb
interactions between the abruptly truncated conduction electrons at the surface
and the ion cores impose a constraint ordering the surface in a well defined
lamellar structure. This constraint can be thought of as an effective hard wall
ionic potential at the surface where the conduction  electron density abruptly terminates.

The existence of surface layering in liquid metals has only recently been
verified unambiguously by experiment, first for liquid Hg\cite{magn95} and
then for liquid Ga.\cite{regan95} Comparison of the temperature
dependence of this layering, however, reveals qualitative differences
between
the two metals.\cite{regan96a,dimasi98} This observation
indicates that details of the interactions such as  the degree of
covalency in the metal may affect the surface structure.

For crystalline metals that can be described by  weak pseudo--potentials, electron--ion scattering
is relatively weak and the non localized or itinerant electrons can be treated
as if they were essentially free.\cite{ashcroft76} Alkali metals with their
nearly parabolic energy bands are the prime example of this class of nearly
free electron (NFE) like metals. In analogy with this, the term NFE metals
applies to liquid metals whose conduction electrons have similar itinerant character and
are only weakly perturbated by a small ionic pseudopotential.\cite{ziman61}
NFE liquid metals can be contrasted with those whose conduction electrons
are partially localized in covalent
    bonds.\cite{selloni95}
Evidence for such covalent
character in liquid metals may be obtained from measurements of the optical
properties as well as from the bulk liquid structure
    factor.\cite{march90}

In fact, both liquid Ga and Hg show substantial deviations from NFE
behavior. In liquid Ga, a shoulder in the bulk liquid structure factor
indicates the presence of directional bonding in the melt.\cite{nart72,bell89,gong93,cicc94}
Liquid Hg displays a pronounced asymmetry in the first peak of the
bulk liquid structure factor, and the optical constants of liquid Hg
deviate
markedly from the predictions of the Drude theory for free
    electrons,\cite{croz72}
which, for example,  adequately describes the optical constants of the alkali
    metals.\cite{mayer63,heyer86}
In order to understand which aspects of
surface layering
may be affected by deviations
from NFE character, it is necessary to
investigate liquid metals with less tendency towards covalent
bonding. The alkali metals would be ideal candidates in this regard.
Unfortunately, the very low surface tension of liquid alkali metals will
result in a large thermally induced surface roughness, making it
extremely difficult to observe the surface layering experimentally. This is
exacerbated by the fact that alkali metals are very reactive in air but
cannot be investigated under Ultra High Vacuum (UHV) conditions due to
their high vapor pressure.
Liquid Indium, which has a high surface tension,
reasonably low melting point and low vapor pressure does not suffer
from these shortcomings. Moreover, it is considered to be NFE like for the
following reasons. First, the  first peak in the bulk structure factor
of liquid In is highly symmetric, indicating the absence of
significant orientational bonding in the
    melt.\cite{orto66}
Second, quantitative analysis of the shape of the bulk liquid structure
factor yields a twelve-fold coordination with an interatomic distance that
decreases with increasing temperature.\cite{ocke66} This is precisely what
is expected for  a liquid
that can be described by an ideal hard sphere model. Third, the
measured optical properties of liquid In agree with the Drude free electron
theory over a large range of wavevectors.\cite{schu57} Even more
interesting, the optical constants of mixtures of liquid Hg and In differ
appreciably from free electron behavior for Hg rich compositions, but
gradually approach free electron behavior with increasing In content.\cite
{hodg67} A main objective of this study is to compare the surface
structure of the more NFE like liquid metal In to that of the more covalent
liquid Ga.

This fundamental interest in the relationship between the surface structure
and electronic properties of liquid metals bears directly on the topic of
surface reactions, which is of considerable practical importance.
 This is especially true
for liquids, where the atoms are mobile, and reactions at the surface can
directly affect alloying, phase formation and other properties of the bulk.
For this reason, surfaces  and reactions at
surfaces play a critical role in process and extractive
metallurgy.\cite{rich74} We previously demonstrated that on exposure to
oxygen the surface of liquid Ga becomes coated with a uniform 5~\AA\ thick
passivating oxide film that protects the underlying bulk phase from further
oxidation.\cite{regan97} We will show below that the oxidation behavior of
liquid In is fundamentally different from that of Ga.

\section{Experimental Details}

The experiments reported here were primarily performed at the beamline X25
at the National Synchrotron Light Source (NSLS). Following a toroidal focussing
mirror, a Ge (220) crystal was used to select the x-ray wavelength
$ \lambda = 0.653$~\AA\ and to tilt the beam downward onto the horizontal
liquid metal surface. The geometry of this liquid spectrometer has been
described
    elsewhere.\cite{regan97b}
Additional data were acquired at
beamline X22B at the NSLS, using a similar scattering geometry and an x-ray
wavelength of 1.24~\AA . The size of the x-ray beam striking the liquid
metal surface was determined by
1~mm~(horizontal)~$\times $~0.1~mm~(vertical) slits
downstream of the monochromator and the
measured signal was normalized to the incoming beam intensity.

The experimental resolution, determined  primarily by slits in
front of the detector, was varied for different experimental
configurations. Specular reflectivity was primarily measured using
a conventional NaI scintillation detector.  Typical slit settings
for measuring this specular reflectivity at X25 were
4~mm~(h)~$\times $~4~mm~(v). At 600~mm from the sample this yields
a $q_z$  resolution of 0.06~\AA $^{-1}$. For comparable resolution
at the longer wavelength at X22B, a vertical detector slit of 8~mm
was used. Small angle off-specular surface diffuse scattering was
measured using a  position sensitive
    detector
(PSD) aligned within the reflection plane.\cite{PSD} For a given
incident angle, the PSD detects both the specular reflection and
off-specular diffuse intensity in the reflection plane up to
$\approx 0.5^{\circ }$  away from the specular peak. We used a
coarse vertical resolution for the PSD in the multi channel mode
(64 channels on 20\,mm). Away from the specular peak, where the
intensity varies slowly with position, data from neighboring
channels were averaged together for improved statistics. Counting
all photons hitting the detector, and disregarding the positional
information, the PSD can be used as a conventional counter. In
this ``single channel'' mode the slit settings were 20~mm
vertically and 3~mm horizontally, corresponding to a $q_z$
resolution of 0.32~\AA $^{-1}$. For all experimental arrangements,
the signal originating at the surface was separated from the bulk
diffuse background by subtracting intensities measured with the
detector moved approximately $\pm $0.4$^{\circ }$ (corresponding
to $\approx 0.67\,{\AA}^{-1}$) transverse to the reflection
plane.\cite{regan97b}

The linearity
of the PSD was checked by comparison of diffuse scans
taken with the PSD  to diffuse data taken  by moving
the scintillation detector within the reflection plane, and by measuring the specular
reflectivity with the stationary  PSD in a single-channel mode. As we will discuss
below, once scaled for the different  resolution, the different types of
scans perfectly overlap each other.

Indium of 99.9995\% purity was contained in a Molybdenum pan within an
Ultra
High Vacuum (UHV) chamber to maintain a clean surface.\cite{XPS} Details of
the UHV chamber design can be found elsewhere.\cite{regan97b} The sample
was heated  above the melting point of
liquid In to $170\pm 5~^{\circ }$C.
After a three day bakeout,
the pressure was
$8\times 10^{-10}$~Torr, with a partial oxygen pressure
$< 8\times 10^{-11}$, for which the formation time  for an oxide
monolayer is  calculated
to be several days.\cite{roth90} Although patches of oxide were present
on the In ingot when it was introduced into the chamber, this  oxide was
removed by Argon ion sputtering (2.5~keV, 20~$\mu $A ion current)
at an Ar partial pressure of about $5\times 10^{-5}$~Torr.
The impact of the ion beam induced
 mass flow along the surface that
transported the oxide patches towards the ion beam, thus cleaning
the entire liquid--vacuum interface and not only the area hit
directly by the Ar$^{+}$
    ions.\cite{fine81}
Data presented here were always
taken within one to two hours of sputter cleaning the sample.
Since
data acquired without sputtering but within a few hours reproduced each
other, and because of the unique growth properties of oxide on liquid In
which  will  be presented below, we are confident that the measured
scattering is intrinsic to a clean liquid Indium surface.

A major improvement over experiments described previously is the
incorporation of active vibration isolation. Previous measurements eliminated
mechanically induced surface waves by utilizing the viscous drag at the
bottom of thin samples. These samples had curved surfaces,
requiring a time consuming technique to measure the
    reflectivity\cite{regan97b} and
making it very difficult to measure at grazing incidence, important for the
detection of in-plane ordering. In this work, the rigid UHV chamber and ion
pump assembly were mounted on an active vibration
    isolation unit,\cite{JRS}
similar to that used previously in our studies of liquid
    Hg.\cite{magn95}
With this system we were able to use a sample pan 5~mm deep and 60~mm in
diameter, resulting in flat samples where the angular deviation of
the surface normal from the vertical across the surface was negligible
compared to the critical angle for total external reflection of x-rays.

\section{Theory}

\begin{figure}[tbp]
\unitlength1cm \epsfig{file=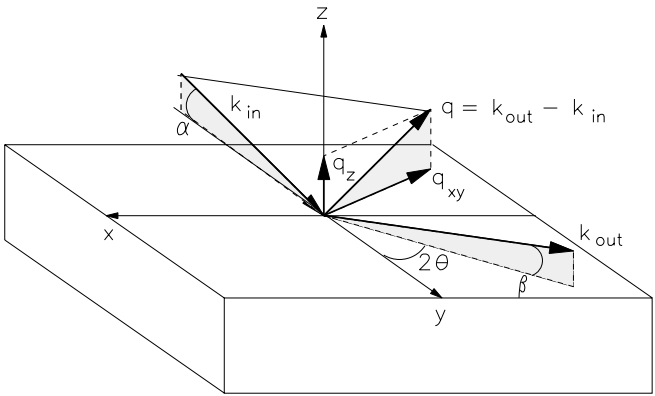, angle=90,
width=1.0\columnwidth} \caption{ Sketch of the geometry of x-ray
scattering from the liquid surface with $\alpha$ and $\beta$
denoting incoming and outgoing angle, the incoming and outgoing
wavevector $\vec{k}_{in}$ and $\vec{k}_{out}$ respectively and the
azimuthal angle $2\theta$. The momentum transfer $\vec{q}$ has an
in-plane component $q_{xy}$ and a surface-normal component $q_z$.
} \label{in_fig:geom}
\end{figure}

The scattering geometry is given in Figure~\ref{in_fig:geom},
defining the liquid surface to be lying in the $x$-$y$ plane, with
x-rays incident at an angle $\alpha$ and collected by the detector
at an elevation angle $\beta$ and azimuthal angle $2\theta$.
The momentum transfer $\vec{q}$ can be decomposed
into surface normal, $q_{z}$, and surface parallel, $q_{xy}$, components
given by:
\begin{equation}
\begin{array}{ll}
    q_{z}=\frac{2\pi }{\lambda }(\sin \beta +\sin \alpha ) \\ \\
$and$ \\ \\
      q_{xy}=\frac{2\pi }{\lambda } \sqrt{\cos ^{2}\alpha +
    \cos ^{2}\beta -2\cos \alpha \cos \beta \cos 2\theta } \: .
    \end{array}
\end{equation}
In the following, we develop a formula for the differential cross section
$d\sigma/d\Omega$ for x-ray scattering from
a rough and structured liquid surface.
The general expression for the  differential cross section
for x-ray scattering from a three dimensional electron distribution
$\rho(\vec{r})$ at a given location $\vec{r}$
 can be expressed in terms of the electron  density correlation function
$\langle \rho(\vec{r}) \rho(\vec{r}\prime)\rangle =
\langle \rho(\vec{r}-\vec{r}\prime) \rho(0)\rangle$:\cite{guinier}
\begin{equation}
\begin{array}{ll}
    \frac{d\sigma }{d\Omega }= V \left( \frac{e^{2}}{mc^{2}}\right)^2
\int d^3 (\vec{r} - \vec{r}\prime)  \langle \rho(\vec{r} -
\vec{r}\prime) \times \\ \\
\times \rho(0)\rangle \exp\left[\imath
\vec{q} \cdot (\vec{r} - \vec{r}\prime) \right] \label{in_general}
\end{array}
\end{equation}
where $V$ is the illuminated volume.
If the  density distribution  is homogeneous within the $x$-$y$ plane but inhomogeneous
normal to the surface, the density correlation function depends on  the different
positions $\vec{r}_{xy} - \vec{r}_{xy}\prime $
on the surface
and the distances $z$ and $z\prime$ from the surface.
For an x-ray beam of a cross sectional
area $A_0$ incident on the surface at an angle $\alpha$ relative to the
$x$-$y$ plane, the illuminated area is $A_0/\sin(\alpha)$
and the differential cross section can be written:\cite{bras88}
\begin{equation}
\begin{array}{ll}
\frac{d\sigma }{d\Omega }=  \frac{A_0}{\sin(\alpha)} \left(
\frac{e^{2}}{mc^{2}}\right)^2 \int dz dz\prime d^2 \vec{r}_{xy}
\langle \rho(\vec{r}_{xy},z) \times \\ \\ \times \rho(0,z\prime)
\rangle \exp\left[\imath q_z (z-z\prime) + \imath \vec{q}_{xy}
\cdot \vec{r}_{xy}  \right] . \label{in_surface}
\end{array}
\end{equation}
For a homogeneous liquid that is sufficiently far away from any critical
region, the only long wavelength excitations that give rise to significant scattering
are thermally excited surface capillary waves. In this case,  there is some
length $\xi$ such that for $|\vec{r}_{xy}| > \xi$ the density density
correlation function can be expressed in terms of a  density profile
$\tilde{\rho} [z - h( \vec{r}_{xy} )]$ defined relative to the local position
of the liquid surface $h ( \vec{r}_{xy} )$
and  capillary wave induced relative variations
of the height, $h(\vec{r}_{xy}) - h (\vec{r}_{xy}\prime) $.\cite{comment1}
\begin{equation}
\label{in_height} \langle \rho(\vec{r}_{xy},z) \rho(0,z\prime)
\rangle  \stackrel{ |\vec{r}_{xy}| > \xi}{\longrightarrow}
\tilde{\rho} (z - [h(\vec{r}_{xy} ) - h (0)] ) \tilde{\rho}
(z\prime)
\end{equation}
Defining $\delta \rho(\vec{r}_{xy},z) = \rho(\vec{r}_{xy},z) -  \tilde{\rho}(z-h(\vec{r}_{xy}))$,
the differential cross section from a liquid surface can  now be written:
\begin{equation}
\begin{array}{ll}
\frac{d\sigma }{d\Omega }=  \frac{A_0}{\sin(\alpha)} \left(
\frac{e^{2}}{mc^{2}}\right)^2 \int dz dz\prime d^2\vec{r}_{xy}
\times \\ \\
\times \exp\left[\imath q_z (z-z\prime)+ \imath \vec{q}_{xy}\cdot
\vec{r}_{xy}\right] \times \\ \\ \times
 \left\{ \tilde{\rho} (z -
[h(\vec{r}_{xy} ) - h (0)] ) \tilde{\rho} (z\prime) + \langle
\delta \rho(\vec{r}_{xy},z) \delta \rho(0,z\prime)\rangle \right\}
\label{in_sepacross}
\end{array}
\end{equation}
The second term in the integrand is only non-vanishing for regions $|\vec{r}_{xy}| \leq \xi $.
This term gives rise to both the
bulk diffuse scattering and surface scattering
at momentum transfers having large $q_{xy}$ components.\cite{bras88}
The effect of this term can be separated experimentally from the specular reflection
and may be omitted in the following discussion of scattering close to the specular condition.
A change of variables yields
\begin{equation}
\begin{array}{ll}
\frac{d\sigma }{d\Omega } \approx  \frac{1}{16\pi^2} \left(
\frac{q_c}{2} \right)^4 \frac{A_0}{\sin(\alpha)} \frac{| \Phi
(q_z)|^2}{q_z^2}\times \\ \\  \times \int_{|\vec{r}_{xy}| > \xi}
d^2 \vec{r}_{xy} \langle \exp \left\{ \imath q_z  h(\vec{r}_{xy})
\right\} \rangle \exp \left[ \imath \vec{q}_{xy} \cdot
\vec{r}_{xy} \right]
\end{array}
\end{equation}
where $q_c$ is the critical angle for total external reflection of x-rays and the
surface structure factor
\begin{equation}
\begin{array}{ll}
\Phi (q_z) = \frac{1}{\rho_{\infty}} \int dz \frac{d\tilde{\rho}
(z) }{dz} \exp(\imath q_z z) = \\ \\
 = \frac{-\imath
q_z}{\rho_{\infty}} \int dz  \tilde{\rho} (z)  \exp(\imath q_z z)
\label{eq:structure}
\end{array}
\end{equation}
is the Fourier transform of the  local or  intrinsic density profile
$ \tilde{\rho} (z) $,
 independent of $\vec{r}_{xy}$.
In this equation, $\rho_{\infty}$ is the bulk electron density.

For the case of thermally excited capillary waves on a liquid surface, the height fluctuations
can be characterized by their statistical average $\langle | h(\vec{r}_{xy}) |^2 \rangle$.\cite{bras88}
Sinha et al. have shown that integration over these height fluctuations yields the following
dependence of the scattering on $q_{xy}$ and $q_z$:\cite{sinha88}
\begin{equation}
\int_{|\vec{r}_{xy}| > \xi} d^2 \vec{r}_{xy}  \langle \exp \left\{
\imath q_z  h(\vec{r}_{xy})  \right\} \rangle \exp \left[ \imath
\vec{q}_{xy} \cdot \vec{r}_{xy} \right] =
\frac{C}{q_{xy}^{2-\eta}} \label{in_sinha}
\end{equation}
with $C$ to be determined and
$$
\eta = \frac{k_BT}{2\pi \gamma} q_z^2 .
$$
The value of $C$ can be determined
by integration over all
 surface capillary modes
having wavevectors smaller than the upper wavevector cutoff
$ q_{max} \equiv \pi/\xi $:
\begin{equation}
\begin{array}{ll}
C = \int_{|\vec{q}_{xy}| < \pi/\xi} d^2 \vec{q}_{xy} \int_{|
\vec{r}_{xy} | > \xi} d^2 \vec{r}_{xy}  \times \\ \\ \times
\langle \exp \left\{ \imath q_z  h(\vec{r}_{xy})  \right\} \rangle
\exp \left[ \imath \vec{q}_{xy} \cdot \vec{r}_{xy} \right]  \approx \\ \\
\approx 4\pi^2 \lim_{ \vec{r}_{xy} \rightarrow 0}
 \langle \exp \left\{ \imath q_z h(\vec{r}_{xy})  \right\} \rangle = 4\pi^2
\end{array}
\end{equation}
Therefore, the properly normalized differential cross section for
scattering of x-rays  from a liquid surface is:

\begin{equation}
    \frac{d\sigma }{d\Omega } = \frac{A_{0}}{\sin \alpha }
    \left( \frac{q_c}{2} \right) ^{4}\frac{k_B T}{16\pi ^2 \gamma }
    \left| \Phi (q_z)\right| ^2
    \frac{1}{q_{xy}^2} \left( \frac{q_{xy}}{q_{\mbox{\scriptsize max}}}
    \right) ^\eta \: .
\label{in_eq:final}
\end{equation}

The intensity measured at a specific scattering vector $\vec{q}$
is obtained by integrating Eq.~\ref{in_eq:final} over the solid
angle $d\Omega$ defined by the detector acceptance of the
experiment. For the specular reflection $q_{xy} = 0$
($\alpha=\beta$ and $2\theta =0$), the integral is centered at
${q}_{xy} = 0$. The projection of the detector resolution onto the
$x$-$y$ plane is rectangular\cite{bras88} and the above mentioned
integration has to be done numerically. However, an analytical
formula can be obtained for the specular reflectivity if the
detector resolution is assumed to be a circle of radius $q_{res}$,
 independent of $q_z$\cite{regan97b}:
\begin{equation}
    \frac{R(q_z)}{R_f(q_z)} = \left| \Phi (q_z) \right| ^2
    \left( \frac{q_{\mbox{\scriptsize res}}}{q_{\mbox{\scriptsize max}}}
    \right) ^\eta = \left| \Phi (q_z) \right| ^2
    \exp [ - \sigma _{\mbox{\scriptsize cw}}^2q_z^2 ]
\label{in_eq:circle}
\end{equation}
where
$$
R_f (q_z) = \left| \frac{q_z - \sqrt{q_z^2 -q_c^2}}{q_z +
\sqrt{q_z^2 -q_c^2}} \right|^2 \approx \left(\frac{q_c}{2
q_z}\right)^4 \ \  \ \ { for}\ \ q_z \stackrel{>}{\sim} 5q_c
$$
is the Fresnel reflectivity from classical optics for a flat ($h(\vec{r}_{xy}) = h (0)$
for all $h(\vec{r}_{xy})$)
and structureless ($|\Phi (q_z)|^2 =1$) surface. The quantity
$\sigma _{\mbox{\scriptsize cw}}$
denotes the capillary wave roughness:
\begin{equation}
\sigma _{\mbox{\scriptsize cw}}^2 = \frac{k_B T}{2\pi \gamma }
    \ln \left(
    \frac{q_{\mbox{\scriptsize res}}}{q_{\mbox{\scriptsize max}}
    }\right) \:  .
\label{in_eq:cw}
\end{equation}
In the experiments presented here, the measured specular
reflectivity was analyzed by numerical integration of the
theoretical model over the rectangular resolution defined by the
detector slits. In practice, the specular reflection is measured
with coarse resolution and the results based on
Eq.~\ref{in_eq:circle} fall within 5\% of the exact values
obtained from integrating over Eq.~\ref{in_eq:final}. On the other
hand, the off specular diffuse scattering is measured with much
finer resolution and near the specular peak, $\vec{q}_{xy} = 0$,
it has to be analyzed by numerical integration of
Eq.~\ref{in_eq:final} over the slit defined resolution. In
addition, numerical integration of Eq.~\ref{in_eq:final} is
necessary to obtain the correct ratio of specular reflectivity to
off-specular diffuse scattering.

In Eq.(\ref{in_eq:structure}) we have introduced a local structure
factor which is the Fourier transform of the local or intrinsic
density profile. An alternative representation is to define a
macroscopically averaged density profile $\langle \rho(z)
\rangle_T$ such that
$$
\frac{1}{\rho_{\infty}}
\int dz \frac{\langle \rho (z)\rangle_T}{dz} \exp(\imath q_z z) = \Phi (q_z) \exp(-1/2 \sigma_{cw}^2 q_z^2)
$$
This macroscopically averaged density profile, $\langle \rho (z) \rangle_T$, can be shown to be equal to
the convolution of the local density profile, $ \tilde{\rho} (z) $, with the associated
Gaussian distribution of height fluctuations characterized by
$$
\sigma_{cw}^2 = \langle \left| h(\vec{r}_{xy}) \right|^2 \rangle = \frac{k_BT}{2\pi\gamma}
\ln \left(\frac{1}{|\vec{r}_{xy}| q_{max}}\right) .
$$
Unfortunately, it is difficult to extract a precisely defined $\langle \rho (z) \rangle_T$ from a single
reflectivity measurement because of the above mentioned problem of the variation of the resolution with
$q_z$.

A second difficulty with reflectivity  measurements, even when $| \Phi (q_z)|^2$ can be separated
from the thermal effects, is that the  direct inversion of $| \Phi (q_z)|^2$ in order to obtain
 $ \tilde{\rho} (z) $  is not possible because the phase information necessary to perform the
Fourier transform in Eq.~\ref{in_eq:structure} is lost in any
intensity measurement. The common practice to overcome this
problem is to assume a model for the density profile, which is
then Fourier transformed and fitted to the experimentally
determined structure factor. In this study, the liquid metal is
modelled by layers of different electron densities parallel to the
surface. The layers represent planes of atoms, separated by a
distance $d$, and having increasing width the further they are
below the surface. Mathematically, the electron density is
described  by  a semi-infinite sum of Gaussians, normalized to the
bulk density $\rho _{\infty }$:
\begin{equation}
     \tilde{\rho} (z) =\rho _\infty \sum _{n=0} ^\infty
    \frac{d/\sigma _n}{\sqrt{2\pi }}
    \exp \left[ -(z-nd)^{2}/2\sigma _n ^2 \right]
    \otimes  F_{\mbox{\scriptsize In}}(z).
\label{in_eq:rho}
\end{equation}
Here $\otimes $ denotes convolution, $F_{\mbox{\scriptsize
In}}(z)$ is the atomic x-ray scattering form factor of In, and
$\sigma _n ^2 = n\bar{\sigma}^2 + \sigma _{0}^{2}$, where
$\bar{\sigma}$ and $\sigma _0$ are constants. This form for
$\sigma _n$ produces a quadratic increase in the Gaussian width
with distance $z$ below the surface, so that the parameter
$\bar{\sigma}$ is related to the decay length for surface layering
and the model density approaches the bulk density $\rho_{\infty}$
for $\sigma_n \gg d$. An advantage of using Gaussian functions to
model $ \tilde{\rho} (z) $ is that an expression for $\langle \rho
(z) \rangle_T$  is obtained from Eq.~\ref{in_eq:rho} by replacing
$\sigma_n$ with $\sigma_n^T$ and defining $(\sigma_n^T)^2 =
\sigma_{cw}^2 + \sigma_n^2$. Note that the effective width of the
individual layers becomes $(\sigma_n^T)^2$ which explicitly
demonstrates the additivity of the different contributions to  the
surface roughness from capillary modes (for which ${q}_{xy} <
q_{max}$) and short wavelength modes (for which ${q}_{xy}  >
q_{max}$) that are incorporated into the model for $ \tilde{\rho}
(z) $.

\section{Experimental Results}

\begin{figure}[tbp]
\unitlength1cm \epsfig{file=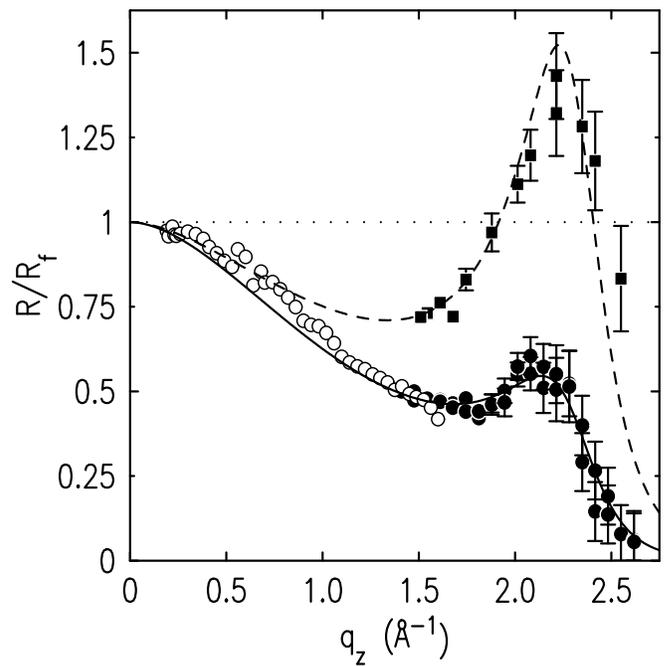, angle=90,
width=1.0\columnwidth} \caption{Specular x-ray reflectivity for
liquid Indium at 170~$^\circ$C taken with vertical detector
resolutions of 0.06\AA$^{-1}$ (scintillation detector; open
circles: X22B, closed circles: X25)
 and 0.32\AA$^{-1}$ (PSD in single channel mode; squares: X25).
The reflectivity is normalized to the Fresnel reflectivity $R_f$
of a flat In surface. Solid lines are fits as described in the
text. Data from X22B was not incorporated in any fits.}
\label{in_fig:refl}
\end{figure}

The specular reflectivity measured from liquid In at
$170\pm 5$~$^\circ$C
demonstrates the presence of surface layering.
    Figure~\ref{in_fig:refl} shows the
reflectivity as a function of $q_z$, normalized to the Fresnel
reflectivity of a sharply terminated In surface with a critical angle
$q_c=0.052$~\AA $^{-1}$. The circles show the data taken with the
scintillation detector, with a $q_z$ resolution of 0.06~\AA $^{-1}$.
Open circles represent the small angle data taken at X22B ($\lambda$=1.24\,{\AA})
whereas the closed circle data have been taken at X25 ($\lambda$=0.65\,{\AA}).

 Filled
squares represent data taken with the PSD in a single channel mode, with
the
vertical resolution coarser by a factor of 5.
The large increase in signal with decreasing resolution immediately demonstrates
the significance of the diffuse scattering.
The reflectivity at small and intermediate wavevectors falls below the Fresnel
value calculated for an ideally flat and abruptly terminated surface. This is
due to the roughness of the liquid In surface which scatters photons out of the
specular condition.
The most prominent feature in these data is the broad peak centered near
2.2~\AA $^{-1}$. This peak is due to constructive interference of x-rays  from
layers ordered parallel  to the surface. This very distinctive structural feature of
liquid metal surfaces has been observed earlier for
    Hg\cite{magn95} and
    Ga.\cite{regan95}
The two data sets show the same features, although as a
consequence of integrating the diffuse scattering over coarser resolution,
the intensity is greater for data taken with the PSD.

The lines in
    Figure~\ref{in_fig:refl}
illustrate fits of the density profile model discussed above to these data.
Three surface layering parameters, $d$, $\overline{\sigma}$, and $\sigma _0$,
determine the form of $\Phi (q_{z})$, which is then used in
    Eq.~\ref{in_eq:final}
along with $T= 170^\circ $C and the experimentally determined
value for the surface tension of liquid In at that temperature, $
\gamma = 0.556$~N/m.\cite{iida93} As discussed above, we assign
the short wavelength cutoff for capillary waves, $q_{max} \approx
\pi/\xi$ , to be $\sim 1 {{\AA}^{-1}}$ where $\xi$ was taken to be
the nearest neighbor atomic distance in the bulk
melt.\cite{comment2} The integral over
    Eq.~\ref{in_eq:final}
is then performed numerically over the known resolution volume.
Surface layering parameters
obtained from the scintillation detector data are
$d=2.69\pm 0.05$~\AA ,
$\overline{\sigma} = 0.54 \pm 0.06$~\AA , and
$\sigma _0 = 0.35 \pm 0.04$~\AA\ (solid line in
    Figure~\ref{in_fig:refl}).
Essentially the same parameters (with larger error due to poorer
statistics) result from a fit of the model to
the PSD data (broken line in
     Figure~\ref{in_fig:refl}).

\begin{figure}[tbp]
\unitlength1cm \epsfig{file=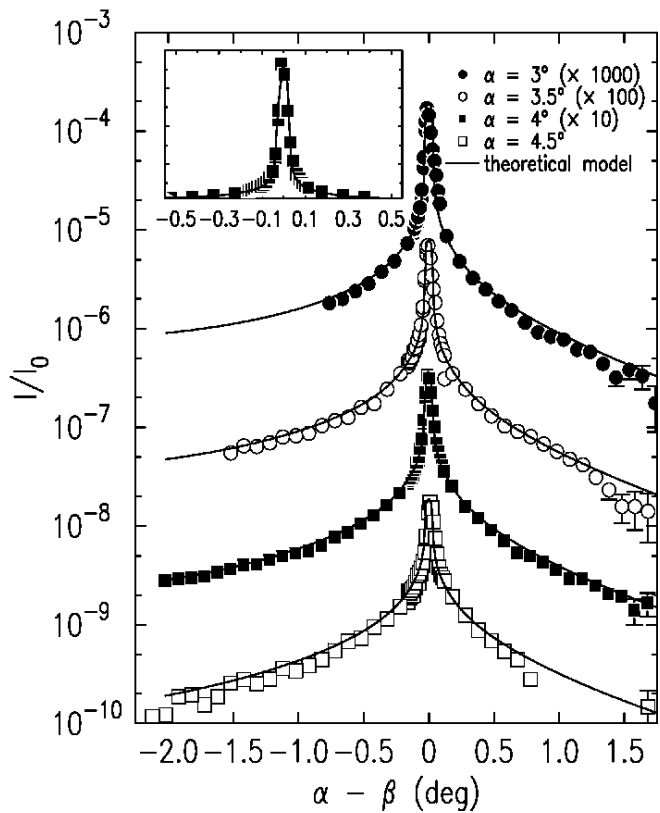, angle=90,
width=1.0\columnwidth} \caption{Diffuse scattering as a function
of scattering angle $\protect\beta$ for different fixed incoming
angles $\protect\alpha$. Solid lines: Diffuse scattering
calculated  from the experimentally determined structure factor
with  no further adjustable parameters. Inset: linear plot
emphasizing the fit near the specular peak for $\protect\alpha =
4.5^{\circ}$%
} \label{in_fig:diff}
\end{figure}

Although the resolution dependence of the reflectivity of liquid metals
provides a test of the thermal capillary wave theory prediction, a
more rigorous test is the measurement of the spectral density of the
off specular diffuse scattering. We have measured diffuse intensity
over a $\beta $ range straddling the specular condition.
Intensities normalized to the direct beam are  shown in
    Figure~\ref{in_fig:diff}
for several choices of $\alpha $. Data at extreme values  of
$(\beta -\alpha) $ are limited by the intense background due to
bulk scattering at large $\beta$ and by the scattering geometry
at small $\beta$. The asymmetry of the wings centered around the specular
ridge ($\beta -\alpha =0)$ arises from the $\beta $ dependence
of both the exponent $\eta $ and the surface structure factor $\Phi (q_z)$.

The solid lines are obtained by calculating the intensity from
    Eq.~\ref{in_eq:final},
integrating numerically over the resolution function
and subtracting a similar calculation for the
background scan taken out of the reflection plane.
The calculation incorporates the surface structure factor $\Phi (q_z)$
determined from the reflectivity measurements along with the other fixed
quantities
($\gamma $, $k_B T$, and $q_{\mbox{\scriptsize max}}$),
without further adjustable parameters. We find excellent agreement between
this model and the experimental data for the entire range studied,
including the wings
    (Figure~\ref{in_fig:diff}) and the specular region (inset of
    Figure~\ref{in_fig:diff}).
The implications of this agreement will be discussed
in the next section.

To study the oxidation properties of liquid Indium
we exposed the In surface to controlled amounts of oxygen through a
bakeable UHV leak valve.
During oxidation, macroscopic oxide clumps, large enough to
be observed by eye, formed  at the edges of the sample, while the
center remained clean. X-ray reflectivity measured during oxidation showed
no changes until the floating oxide regions grew large enough to reach the
area illuminated by the x-ray beam, at which point the macroscopically rough surface
scattered  the reflected signal away from the specular condition.
This result is in sharp contrast to the
formation of the  highly uniform, 5~\AA\  thick, passivating oxide layer
we previously observed to form on liquid Ga under the influence
of the same amount of oxygen.\cite{regan97}

\section{Discussion}

\begin{figure}[tbp]
\epsfig{file=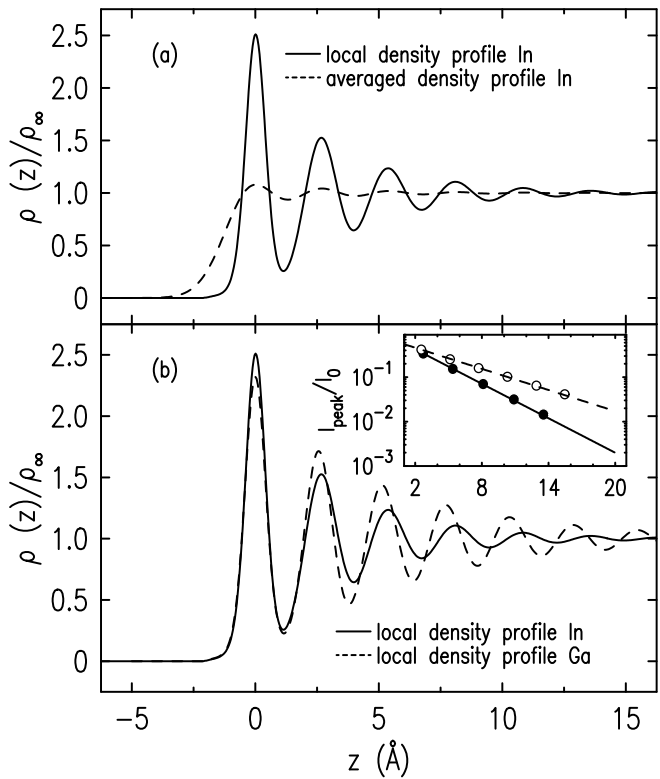, angle=90, width=1.0\columnwidth}
\caption{ (a) Comparison of the local real space density profile
for liquid Indium (---) with the thermally averaged density
profile for the same metal (- - -). The averaged density profile
is directly accessible by experiment and has been measured at
170$^{\circ}$\,C. (b)  Real space local density profile for liquid
Gallium (- - -) and
liquid Indium (---) after the removal of thermal broadening. Densities $%
\protect\tilde{\rho} (z)$ are normalized to the bulk densities
$\protect\rho_{\infty}$ of liquid Gallium and Indium. Inset: Decay
of the amplitude of the surface--normal density profile for liquid
Ga (open circles) and liquid In (filled circles). The lines
represent the fit of this decay of the surface layering to the
form $\exp(-z/l)$ for liquid Ga (---) and In (- - -). }
\label{in_fig:rho}
\end{figure}

The macroscopic density profile extracted from the reflectivity
measurements is resolution dependent since the density
oscillations are smeared out by thermally activated capillary
waves as described in the Theory section. This smearing depends on
the temperature and surface tension and hence also varies for
different liquids. The  macroscopic density profile directly
extracted from the experiment is the  averaged density profile
$\langle \rho (z)\rangle_T$  shown in previous
publications.\cite{magn95,regan95,regan96a,regan97b} In order to
compare the intrinsic layering properties in different liquid
metals, it is necessary to remove these thermal effects that vary
from metal to metal and experiment to experiment and to obtain the
intrinsic or  local density profile $ \tilde{\rho} (z)$. Previous
temperature dependent reflectivity measurements  of liquid Ga
around the specular position have shown that the resolution
dependent roughness is well described by the capillary wave (CW)
prediction.\cite{regan96a} Similar conclusions were reached from
$T$-dependent measurements of liquid paraffins.\cite{ocko94} Here,
we employed a different approach  to verify that  the CW
prediction holds for liquid In, performing  diffuse scattering
measurements at a single temperature at different angles of
incidence. The magnitude and the angular dependence of the diffuse
scattering is  in full compliance with the predictions of the
capillary wave model without
 using any adjustable parameters. This is demonstrated by the perfect agreement
of the diffuse scattering profiles with the theoretical curves in
    Figure~\ref{in_fig:diff}. This allows us to directly calculate the local
density profile $ \tilde{\rho} (z) $ by simply setting $\eta =0$.
A comparison between the local density profile $ \tilde{\rho} (z)$
and the temperature averaged density profile $\langle \rho (z)
\rangle_T$  is made in Figure~\ref{in_fig:rho}(a).

In this
single-$T$ approach the effects of the intrinsic, $T=0$, roughness
$\sigma _0$ and the cutoff $q_{\mbox{\scriptsize max}}$ cannot
be unambiguously
    separated,\cite{gelf90}
as was possible for the $T$-dependent measurements in
    alkanes\cite{ocko94}
and     Ga.\cite{regan96a}
However, this does not affect our main conclusion that the surface
roughness, as probed by the DS, is entirely due to thermally activated
capillary waves.

The local density profile of liquid In is compared with that of Ga
in Figure~\ref{in_fig:rho}(b). Profiles derived from the extremal
parameters of fits to $R(q_z)$ and from extremal values for
temperature, surface tension and resolution function are
indistinguishable on the scale of the figure. The amplitude of the
first density peak for In is comparable to that of Ga. This
reflects the fact that in the region nearest the surface, the
density oscillations are determined primarily by $\sigma _0$,
which has similar values for the two metals. Further towards the
bulk, layering is seen to decay faster for In than for Ga. We
quantify this observation by the excellent fit of  the peak
amplitude of the electron density profile to the form $\exp (-z/l
)$. The decay lengths $l =3.5$\,{\AA}$\pm 0.6$ for In and $l
=5.5$\,{\AA}$\pm 0.4$ for Ga differ by an amount well outside  of
the  experimental error. This is illustrated in the inset of
Figure~\ref{in_fig:rho}(b), a semilogarithmic plot of the maxima
of the density oscillations as a function of distance from the
first surface layer. The lines represent the fit of the peak
amplitude to the exponential form and the slope represents the
decay length.

\begin{figure}[tbp]
\epsfig{file=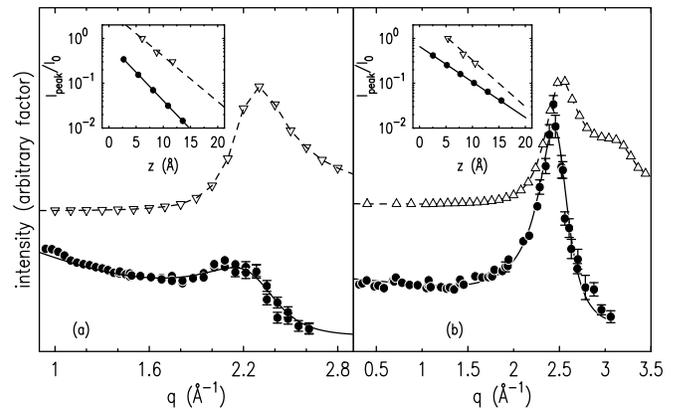, angle=90, width=1.0\columnwidth}
\caption{(a) Bulk ($\triangle$) and surface ($\bullet$) structure
factors for liquid In. The bulk structure factor data are taken
from Orton et al.\protect\cite{orto66} The solid line is a fit of
the model explained in the text to the experimentally determined
surface structure factor. The broken line is a guide for the eye.
The inset compares the decay of the longitudinal surface density
oscillations (filled circles; solid line represents fit to an
exponential) to the decay of the bulk pair correlation function
(open triangles; broken line represents fit to exponential). The
coordinate $z$ of the x-axis represents the distance from the
surface in the case of the decaying surface layering and the
radial distance from a reference atom in the bulk liquid in the
case of the decaying bulk pair correlation.
 The data for the bulk pair correlation
function are taken from Waseda.\protect\cite{waseda80} (b) Bulk
($\triangle$) and surface ($\bullet$) structure factors for liquid
Ga. The bulk structure factor data are taken from Narten et
al.\protect\cite{nart72} Solid lines are fits to reflectivity
data, broken lines are guides for the eye. The inset compares the
decay of the longitudinal surface density oscillations (filled
circles; solid line represents fit to exponential) to the decay of
the bulk pair correlation function (open triangles; broken line
represents fit to exponential). The data for the bulk pair
correlation function are taken from Waseda.\protect\cite{waseda80}
}
 \label{in_fig:bulk}
\end{figure}

It is also instructive to compare the surface structure factor
measured in the reflectivity experiments to the bulk structure
factor measured by standard x-ray or neutron diffraction
(Figure~\ref{in_fig:bulk}(a): Indium; Figure~\ref{in_fig:bulk}(b):
Gallium). The widths of the peaks shown are  inversely
proportional to the decay lengths of the corresponding correlation
functions: the pair correlation function $g(r)$\cite{waseda80} of
the bulk and the layered density profile $ \tilde{\rho} (z)  $  of
the surface. Plotting the amplitudes  of $g(r)$ and $ \tilde{\rho}
(z)  $ on a semi-log scale (inset of  Figure~\ref{in_fig:bulk})
shows that they lie on straight lines. This indicates that the
correlations decay exponentially, and the decay lengths are
obtained as the  negative inverses of the slopes. The inset plots
show clearly that while the surface and bulk decay lengths are
about the same for In, in Gallium the surface induced layering
decays much more slowly than the bulk pair correlation function.
This is consistent  with the NFE nature of In. For Ga, by
contrast, the tendency towards covalency may disrupt the hard
sphere packing and enable surface correlations that differ from
those in the bulk.

Another piece of evidence suggesting that directional bonding in
the melt might have an effect on surface induced layering stems
from the ratio between the surface layer spacing and the bulk
nearest neighbor distance. This nearest neighbor distance is taken
from the analysis of the pair correlation function which is the
Fourier transform of the structure factor depicted in
Figure~\ref{in_fig:bulk}. Due to truncation problems and related
subtleties with the  data analysis, the  bulk nearest neighbor
distance cannot be directly inferred from the maximum in the bulk
liquid structure factor and we use the nearest neighbor distance
given by Iida and Guthrie.\cite{iida93} For liquid In, this ratio
of surface layer spacing to bulk nearest neighbor distance is
2.69{\AA}/3.14{\AA} = 0.86, close to the value of $\sqrt{2/3}
\approx 0.82$ for the hard sphere packing that would be expected
for a NFE liquid metal.
 For liquid Ga, this ratio
is larger, 2.56{\AA}/2.78{\AA} = 0.92. In this case, formation of
directional bonds, eventually leading to $ Ga_2$ dimers, may
prevent close packing.

Tomagnini et al.  recently
considered a possible relationship between surface induced layering in LM and the
stability of crystal facets at metal surfaces.
Although premelting of crystalline interfaces is quite common
in non-metallic crystals,  most metals have at least one close packed face
which does not premelt,
remaining solid up to the melting temperature\cite{toma98}. Tomagnini et al.
use Molecular Dynamics simulations to demonstrate
that when the period of the surface induced layering at the liquid surface,
which they assumed to be $2\pi/q_0$  where $q_0$ is the position for the peak in the bulk liquid
structure factor,
is commensurate with the distance between lattice planes along the normal
to the crystal facet,
this particular facet is stabilized and resistant to premelting.
It would be interesting to correlate the premelting properties of Ga and In crystals
with our results for the period of the surface layering and in particular, to examine
if the effect of directional bonding that affects the surface structure in the liquid
affects premelting of the solid as well.


Our diffuse scattering measurements show that the roughness of the
liquid In surface can be attributed entirely to thermally excited
capillary waves. This indicates that there are no other detectable
inhomogeneities on the surface, such as might be caused by
microscopic oxide patches. This observation is in concert with
independent experiments on organic Langmuir monolayers on water.
If the monolayer is homogeneous\cite{masa,gour97}, the DS can be
described by Eq.(\ref{in_eq:final}) with no adjustable parameters
and no excess scattering is observed --- just as  is the case for
metallic liquid In. On the other hand, if the organic monolayer is
compressed beyond its elasticity limit and becomes inhomogeneous,
Eq.(\ref{in_eq:final}) no longer describes the experimentally
determined DS, and excess scattering is observed  that must
originate from sources other than thermally activated capillary
waves.\cite{masa} This demonstrates the viability of diffuse
scattering as a tool for  studying surface inhomogeneities. This
is of particular interest for investigations on liquid alloys
where --- depending on the type of alloy --- a rich surface
behaviour is expected, ranging from concentration fluctuations for
alloys displaying surface segregation to critical fluctuations for
alloys with a critical consolute
    point.\cite{tost98}

Controlled oxidation of the liquid metal surfaces reveals a further
striking
difference between Ga and In. Although exposure of liquid Ga to air or to a
large amount of oxygen leads to the formation of a macroscopically
thick  and rough oxide film, exposure to controlled amounts of oxygen under
UHV conditions produces a well defined  uniform 5~\AA\ thick oxide
    film.\cite{regan97}
This film protects the underlying bulk phase to a certain
extent from further corrosion, as further dosage at low oxygen pressures
($<2 \times 10^{-4}$~Torr)
was found to have no effect on the oxide thickness or
coverage fraction. This pasivating of the surface  is rather unusual for solid metals, Al
being prominent among the few examples. Liquid In, unlike Ga, was not found
to form a passivating oxide film. Instead, macroscopic clumps of oxide formed
which did not wet the clean In surface.
This corrosion mechanism is rather common for solid metal surfaces,
the oxidation of Fe being the best known example. It is not clear to what
extent this fundamental difference in the mechanism for corrosion is
related to differences in the  structure  of the liquid surface and to what extent to the chemical
affinity for oxidation of the respective liquid metal.

\section{Summary}

We have measured the x-ray reflectivity and small angle off specular
surface
diffuse scattering from liquid In at $170^\circ $C. Our results can be
quantitatively explained by the convolution of thermally excited  surface waves and a
temperature independent surface structure factor, corresponding to
theoretically predicted surface layering. The absence of excess diffuse
scattering beyond that due to thermally excited capillary waves
demonstrates that the liquid--vapor interface is homogeneous in the
surface-parallel direction. The intrinsic layering profile of liquid In,
obtained by removing the capillary wave roughening, is compared to that
previously reported for liquid Ga. For Ga, surface layering
persists  farther  into the bulk than  is the case for
In. This may be attributed to directional bonding in the Ga bond stabilizing
surface induced layering over a larger distance than  is the case for liquid In.
Further evidence suggesting a correlation between the degree of covalency in
the melt and the surface structure stems from the observation that the compression
of the surface layer spacing relative to the bulk nearest neighbor distance
is close to the behavior expected for an ideal hard sphere liquid for In but
considerably different for Ga.
Controlled oxidation of liquid In results in the
formation of a macroscopic  rough oxide, unlike the passivating
microscopic  uniform oxide film that forms on liquid Ga.

\section{Acknowledgements}

The authors thank Masafumi Fukuto for helpful discussions. This work is
supported by the U.S.~DOE Grant No. DE-FG02-88-ER45379, the National
Science
Foundation Grant No. DMR-94-00396 and the U.S.--Israel Binational Science
Foundation, Jerusalem. Brookhaven National Laboratory is supported by U.S.
DOE Contract No. DE-AC02-98CH10886. HT acknowledges support from the
Deutsche Forschungsgemeinschaft.



\begin{thebibliography}{1}
\bibitem{comment1} The microscopic definition of the local density profile $\tilde{\rho}(z-h(\vec{r}_{xy}))$
requires averaging over density fluctuations with wavelengths shorter than $\xi$. It follows
that fluctuations with surface wavevectors $q_{xy} \leq \pi/\xi$ are contained in the definiton
of $\tilde{\rho}(z-h(\vec{r}_{xy}))$.
\bibitem{comment2} This cutoff is selected such that the number of capillary modes is
equal to the number of surface atoms. Choice of a larger cutoff length would imply that
the local density profile includes contributions from averaging over the capillary waves
with the shortest wavelengths.
First principle arguments that might support such a larger cutoff length are beyond the scope of this
paper. A smaller cutoff length on the other hand would imply capillary waves with wavelengths
smaller than the particle size which is not physically reasonable.
\end{thebibliography}
\end{document}